\documentclass[aps,prl,superscriptaddress,twocolumn]{revtex4-2}

\usepackage{graphicx}
\usepackage{dcolumn}
\usepackage{bm}
\makeatletter

\begin{document}
	

\title{Dimensionality-confined superconductivity within SrNbO$_3$-SrTiO$_3$ heterostructures}



\affiliation{Beijing National Laboratory for Condensed Matter Physics, Institute of physics, Chinese Academy of Sciences, Beijing, 100190, China}
\affiliation{School of Physics, University of Chinese Academy of Sciences, Beijing, 100049, China}
\affiliation{Songshan Lake Materials Laboratory, Dongguan, Guangdong, 523808, China}
\affiliation{National Center for Electron Microscopy in Beijing and School of Materials Science and Engineering, Tsinghua University, Beijing 100084, China}

\author{Haoran Wei}
\thanks{These two authors contributed equally.}
\affiliation{Beijing National Laboratory for Condensed Matter Physics, Institute of physics, Chinese Academy of Sciences, Beijing, 100190, China}
\affiliation{School of Physics, University of Chinese Academy of Sciences, Beijing, 100049, China}

\author{Shengru Chen}
\thanks{These two authors contributed equally.}
\affiliation{Beijing National Laboratory for Condensed Matter Physics, Institute of physics, Chinese Academy of Sciences, Beijing, 100190, China}
\affiliation{School of Physics, University of Chinese Academy of Sciences, Beijing, 100049, China}

\author{Yuting Zou}
\affiliation{Beijing National Laboratory for Condensed Matter Physics, Institute of physics, Chinese Academy of Sciences, Beijing, 100190, China}
\affiliation{School of Physics, University of Chinese Academy of Sciences, Beijing, 100049, China}

\author{Yuxin Wang}
\affiliation{Beijing National Laboratory for Condensed Matter Physics, Institute of physics, Chinese Academy of Sciences, Beijing, 100190, China}
\affiliation{School of Physics, University of Chinese Academy of Sciences, Beijing, 100049, China}

\author{Meng Yang}
\affiliation{Beijing National Laboratory for Condensed Matter Physics, Institute of physics, Chinese Academy of Sciences, Beijing, 100190, China}
\affiliation{School of Physics, University of Chinese Academy of Sciences, Beijing, 100049, China}

\author{Qinghua Zhang}
\affiliation{Beijing National Laboratory for Condensed Matter Physics, Institute of physics, Chinese Academy of Sciences, Beijing, 100190, China}
\affiliation{School of Physics, University of Chinese Academy of Sciences, Beijing, 100049, China}
\affiliation{Songshan Lake Materials Laboratory, Dongguan, Guangdong, 523808, China}

\author{Lin Gu}
\affiliation{National Center for Electron Microscopy in Beijing and School of Materials Science and Engineering, Tsinghua University, Beijing 100084, China}

\author{Kun Jiang}
\affiliation{Beijing National Laboratory for Condensed Matter Physics, Institute of physics, Chinese Academy of Sciences, Beijing, 100190, China}
\affiliation{School of Physics, University of Chinese Academy of Sciences, Beijing, 100049, China}
\affiliation{Songshan Lake Materials Laboratory, Dongguan, Guangdong, 523808, China}

\author{Er-Jia Guo}
\email{ejguo@iphy.ac.cn}
\affiliation{Beijing National Laboratory for Condensed Matter Physics, Institute of physics, Chinese Academy of Sciences, Beijing, 100190, China}
\affiliation{School of Physics, University of Chinese Academy of Sciences, Beijing, 100049, China}
\affiliation{Songshan Lake Materials Laboratory, Dongguan, Guangdong, 523808, China}

\author{Zhi Gang Cheng}
\email{zgcheng@iphy.ac.cn}
\affiliation{Beijing National Laboratory for Condensed Matter Physics, Institute of physics, Chinese Academy of Sciences, Beijing, 100190, China}
\affiliation{School of Physics, University of Chinese Academy of Sciences, Beijing, 100049, China}
\affiliation{Songshan Lake Materials Laboratory, Dongguan, Guangdong, 523808, China}



\date{\today}

\begin{abstract}
Interfaces between transition-metal oxides are able to host two-dimensional electron gases (2DEGs) and exhibit exotic quantum phenomena. Here we report the observation of superconductivity below 230 mK for the heterostructure composed of SrNbO$_3$ (SNO) and SrTiO$_3$ (STO). Different from some other counterparts with two insulators, the metallic SNO provides a novel mechanism to form a quasi 2DEG by charge transfer from bulk towards interface under strain. The superconductivity – residing within the strained SNO layer near the interface – is contributed by an electron system with record-low carrier density. Notably, although embedded in a normal metallic layer with a carrier density 4 - 5 orders higher, the electron system is still uniquely well-protected to retain high mobility and lies deep in extreme quantum regime.
\end{abstract}


\maketitle


Interfaces between transition-metal oxides (TMOs) have attracted remarkable interests because their potential in hosting distinct band structures and exotic properties. The first realization of two-dimensional electron gases (2DEGs) at the interface between LaAlO$_3$ (LAO) and SrTiO$_3$(STO) \cite{ref1}  provides a ground-breaking platform to demonstrate unconventional quantum phenomena \cite{ref2,ref3,ref4,ref5,ref6,ref7,ref8,ref9,ref10} among which superconductivity is the most intriguing one because of its importance in fundamental sciences and practical applications \cite{ref11,ref12}. Recently, substitution of STO by KTaO$_3$ (KTO) is expected to introduce stronger topological nature to the superconductivity because KTO’s spin-orbital coupling (SOC) strength is more than 20 times larger \cite{ref13}. Interfacial superconductivity has been realized for heterostructures of EuO/KTO and LAO/KTO with specific crystalline orientations \cite{ref14,ref15,ref16,ref17,ref18,ref19,ref20}.

All observed superconductivity at heterointerfaces between TMOs resides at the side of STO or KTO so far. Thanks to the inversion symmetry breaking, itinerant electron systems can be generated by mismatches in either lattices (strain) or electronic (polar discontinuity) structures. In this sense, dimensionality-confinement plays a crucial role in stabilizing the electron system by creating potential wells. However, the formation mechanism of the electron systems is obscure and hard to clarify because of the complexity and uncertainties for possible origins. Besides interfacial charge-transfer due to polar catastrophe, oxygen vacancies are also possible sources of extra charge carriers \cite{ref21} (see Fig. 1A). Their existence is nevertheless unstable and difficult to quantitatively controlled or analyzed. On the other hand, introduction and adjustment of exotic properties, such as topological band structures and magnetic orders, are not always convenient since the electron system lies in STO or KTO. These properties can only be induced by the adjacent layer via proximation, which is usually difficult and limited by multiple factors such as robustness of the property, location of the electron system, quality of interfaces, etc. Novel methods of constructing and engineering the electron system and its superconductivity is therefore needed.
The difficulty mentioned above may be avoided if the 2DEG lies on the other side of the interface, for which case the behavior of itinerant electrons is directly influenced by the hosting layer. This may be realized by depositing a metallic layer due to its large electron affinity. With similar structure as STO, SrNbO$_3$ (SNO) is metallic perovskite. Nb’s 4d-orbitals is expected to enrich the electron system with relatively strong correlations and large SOC simultaneously, which may lead to interesting phenomena \cite{ref22}. Given the reported success in preparing high quality SNO thin film by pulsed-laser-deposition (PLD) technique \cite{ref23,ref24}, it is an appropriate candidate to realize a 2DEG near the interface, and to explore possible superconductivity and topological behaviors\cite{ref23,ref25}.

Here, we report on the observation of superconductivity in the SNO/STO heterostructure. The transition temperature, between 150 and 230 mK, was found to be dependent on the thickness of SNO and sensitively modulated by strains. Distinguished from any other counterparts, the quasi-two-dimensional (2D) superconducting electron system is uniquely formed by charge transfer from bulk SNO towards the interface driven by strain gradient (see Fig. 1A). Interestingly, it is featured with the record-low carrier density for superconductivity at TMO heterointerfaces as suggested by the quantum oscillations associated with the same electron system. Evidence of non-trivial Berry phases are also observed.

\begin{figure*}
	\centering
	\includegraphics[width=1\linewidth]{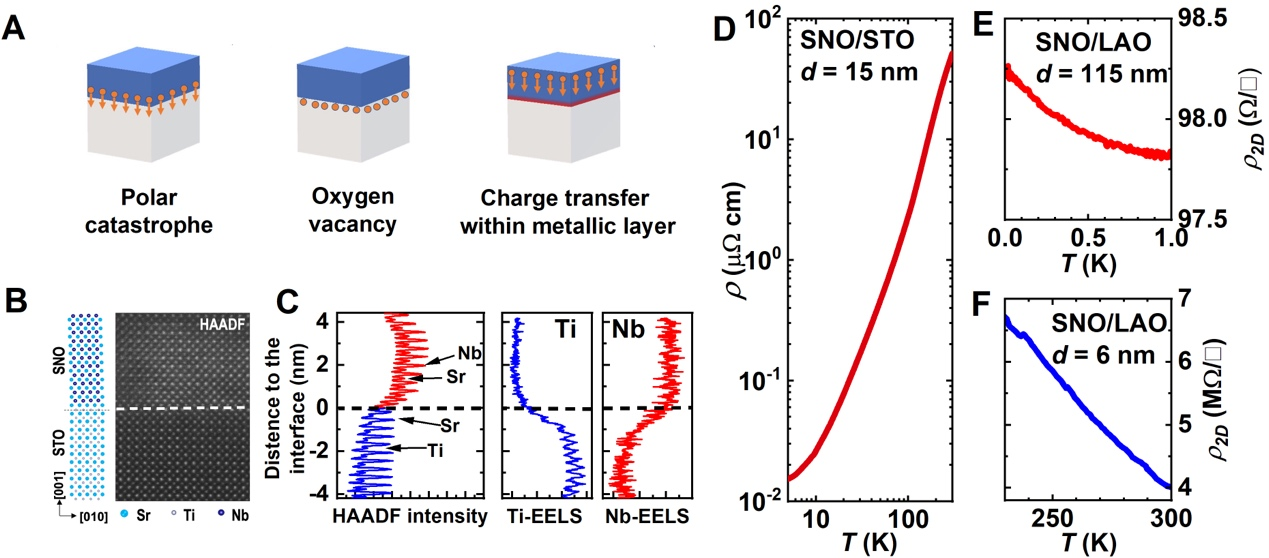}
	\caption{\textbf{(A)} Three possible mechanisms of 2DEG formation at interfaces of TMOs. \textbf{(B)} STEM picture of the SNO/STO interface looking along the [100] direction. \textbf{(C)} Characterizations of chemical intermixing at the interface by HAADF and EELS intensities. \textbf{(D)} Temperature dependence of a SNO/STO heterostructure with $d=$ 15 nm from 300 to 4 K. Temperature dependencies of SNO/LAO heterostructures are shown in \textbf{(E)} and \textbf{(F)} with $d=$ 115 and 6 nm, respectively. The 115 nm sample does not exhibit superconductivity down to 20 mK.}
	\label{Fig1}
\end{figure*}

High-quality SNO films were deposited on (001)-oriented TiO$_2$-terminated STO single-crystalline substrates by PLD technique. An atomically sharp interface could be confirmed by high-angle annular dark-field imaging scanning transmission electron microscopy (HAADF-STEM) (see Fig. 1B). With the lattice constants 3.905 Å for STO and 4.023 Å for SNO, the substrate exerts an in-plane biaxially compressive strain $\epsilon=-3.02\%$. Electron energy loss spectroscopy (EELS) measurements (see Fig. 1C) display the distributions of Ti and Nb elements, suggesting a sharp transition across the interface. The chemical intermixing could not be detected within the measurement resolution (about two unit-cells at maximum) and the HAADF intensity near the interface exhibits the clear periodic arrangement of SrO, NbO$_2$, and TiO$_2$ layers. Both HAADF and EELS results convincingly confirm high crystalline quality of SNO and STO and the sharpness of the interface. A series of SNO/STO heterostructures were prepared with the thickness of SNO layer $d=$ 4, 6, 15, 37, 75, 150 nm, all square-shaped with dimensions of $5\times5$ mm$^2$.

\begin{figure}[b]
	\centering
	\includegraphics[width=1\linewidth]{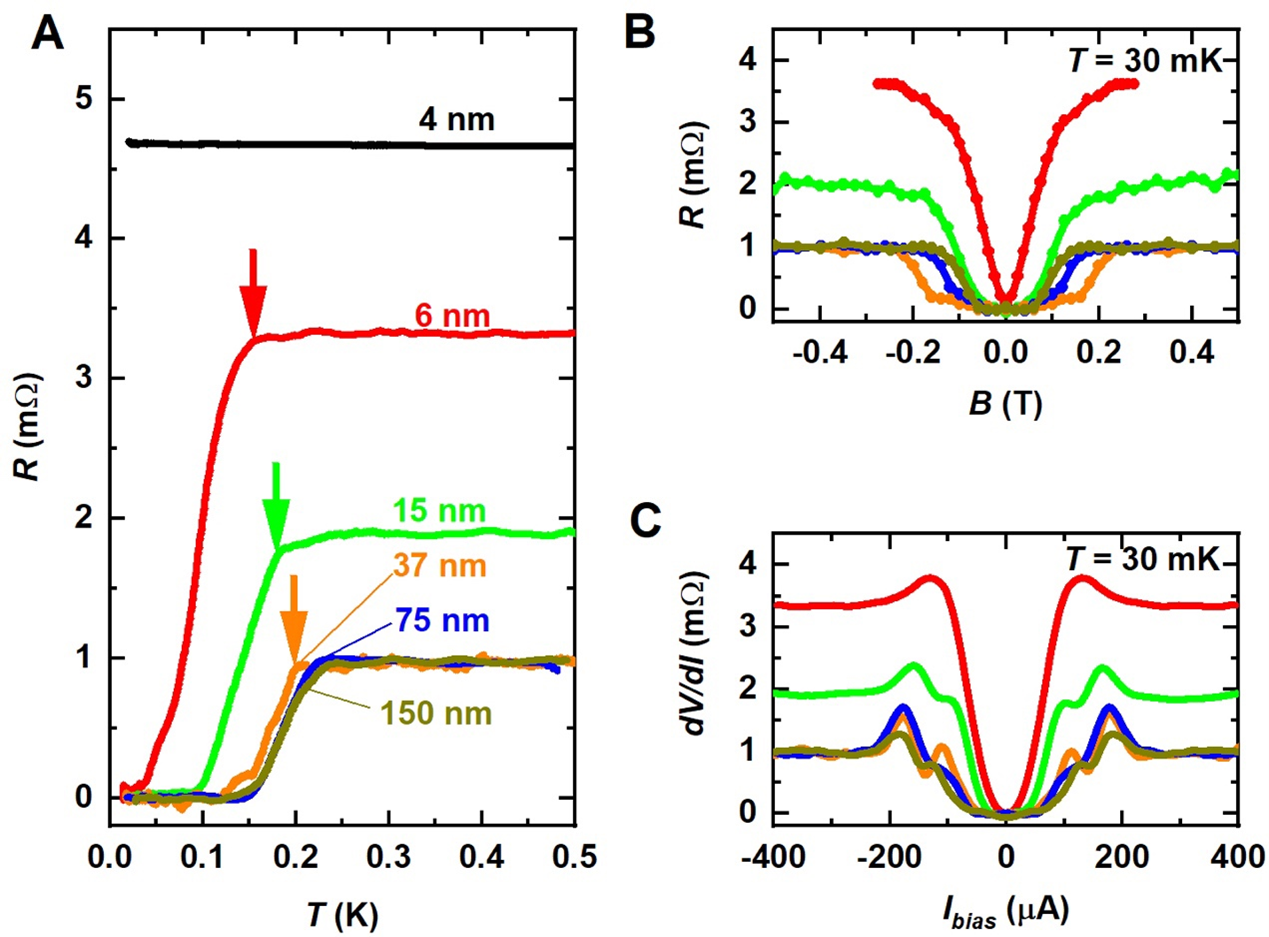}
	\caption{\textbf{(A)} Temperature dependence of resistances for all the SNO/STO samples. The arrows label the superconducting transition temperature $T_c$. \textbf{(B)} Resistances as magnetic field varies between $\pm$ 0.5 T at 30 mK. \textbf{(C)} Differential resistances as bias current $I_{bias}$ varies between $\pm 400$ $\mu$A at 30 mK.}
	\label{Fig2}
\end{figure}

Electric transport measurements were conducted using van der Pauw (vdP) method from room temperatures down to ultralow temperatures ($\sim$20 mK). A typical residual resistance ratio (RRR) is as high as around 3000 as obtained for all sample, providing a further confirmation of high crystallinity. Resistance at low temperatures, e.g. at 0.5 K, monotonically decreases as $d$ rises from 4 to 37 nm but saturates for further increase (see Fig. 2A). Interestingly, superconductivity was observed for all samples with $d\geq 6$ nm. The transition temperature (T$_c$), determined by onset of resistance decrease, monotonically increases with $d$ from 150 to 230 mK. To elucidate the origin of the superconductivity, we prepared another two heterostructures by depositing SNO, with thicknesses of 6 and 115 nm, on substrates of LaAlO$_3$ (LAO). The 6 nm SNO films demonstrates a clear insulating behavior with the resistivity on the order of M$\Omega$ (see Fig. 1E). In comparison, the 115 nm sample, although with a larger conductivity because of strain relaxation, still exhibits a mild insulating behavior below 1.0 K. Despite that, no superconducting transition was observed down to 20 mK (see Fig. 1F). It proves that bulk SNO is non-superconducting, and the observed superconductivity of SNO/STO heterostructures resides near the interfaces.

\begin{figure*}
	\centering
	\includegraphics[width=1\linewidth]{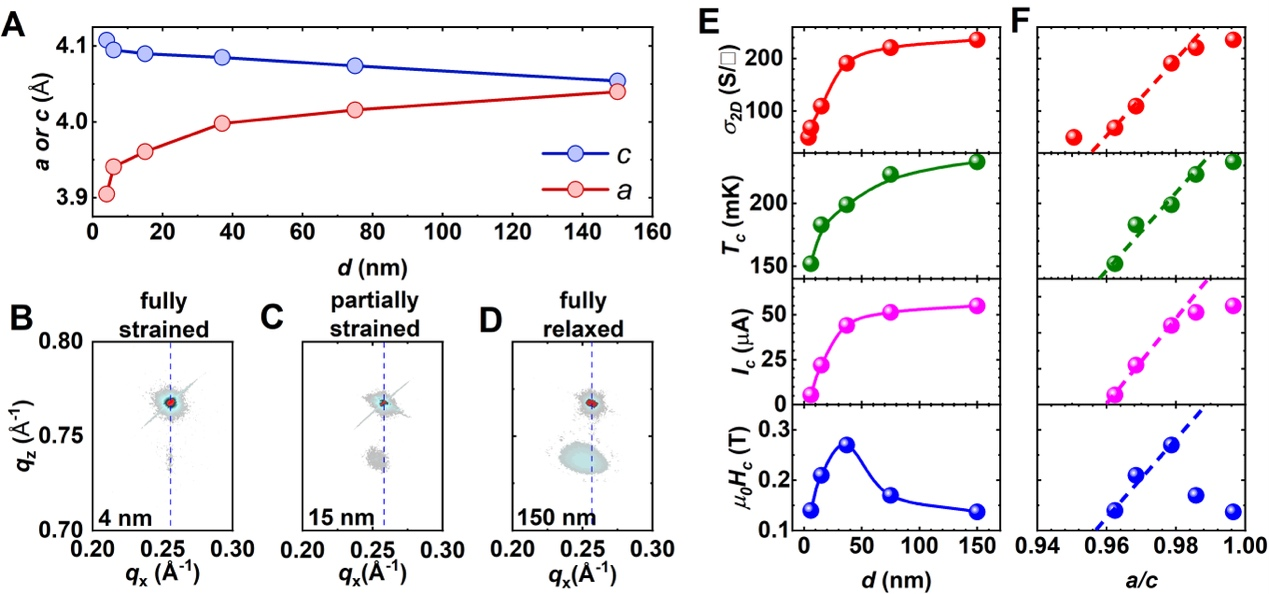}
	\caption{\textbf{(A)} In-plane ($a$) and out-of-plane ($c$) lattice constants as functions of SNO thickness measured by XRD. \textbf{(B-D)} RSM measurements of the films with $d=$ 4, 15, and 150 nm. \textbf{(E)} Thickness dependence of $\sigma_{2D}$, $T_c$, $I_c$, and $H_c$. \textbf{(F)} Dependence on the ratio $a/c$ for the same parameters shown in \textbf{(E)}. The dashed lines indicate linear dependencies.}
	\label{Fig3}
\end{figure*}

We further explored superconducting transitions driven by magnetic field and bias current. The R-B curve shown in Fig. 2B evolves from “V” to “U” shape as d increases. A similar evolution is also observed for the $dV/dI$-I$_{bias}$ curve as shown in Fig. 2C, both suggesting that the superconductivity becomes more robust. The transitions can be parametrized by the critical field $H_c$ – determined by the field at which resistance drop to 90\% of its value at normal state, and the critical current $I_c$ – determined by the onset of sudden increases of $dV/dI$. Fig. 3E summarizes the thickness dependence of two-dimensional conductivity ($\sigma_2D$), $T_c$, $I_c$, and $H_c$, all rapidly increasing before d reaches 37 nm. Moreover, the first three parameters universally tend to saturate while $H_c$ significantly decreases for further increase of $d$.

The universal thickness dependence as described above is likely associated with the in-plane biaxial strain exerted by the substrates. Compressive strain causes decrease of in-plane lattice constant ($a$) and increase of out-of-plane lattice constant ($c$). As $d$ increases, $a$ and $c$ gradually merge due to strain relaxation. Their thickness-dependencies were characterized by comprehensive x-ray diffraction (XRD) measurements (see Fig. 3A). Impressively, the change of lattice constants is also rapid for $d<$ 37 nm and mild for larger $d$, in agreement with the universal thickness dependence. More details of strain relaxation can be seen by reciprocal space mapping (RSM) measurements (see Figs. 3B-3D and Figs. S2 in Supplementary Materials), which demonstrates that partial relaxation of strain takes place within the layer of $d$ between 6 and 37 nm, leading to a vertical strain gradient in this part. The similarity of thickness dependence for both strain and superconducting parameters confirms the strong modulation of superconductivity by misfit strain, which is furtherly supported by the universal linear dependence on the $a/c$ ratio as shown in Fig. 3F.

Superconductivity at insulating TMO interfaces can usually be efficiently tuned by gate voltage. However, field-effect tuning does not apply to the case of SNO/STO because of the screening of itinerant electrons in the SNO layer. The effective modulation by strain must originate from an alternative mechanism. It is highly valuable to explore the underlying mechanisms which may unveil the origin of the superconductivity. It is well-known that band structure could be modulated by strains. For perovskite compounds, the split between $t_{2g}$ and $e_g$ bands is caused by crystal field and therefore sensitive to strains; energies of different orbitals also variate when spatial symmetry is broken. Therefore, strain can possibly modulate carrier densities and band structures. A Dirac point not far from Fermi level, for instance, is predicted to emerge when compressive strains and rotations of NbO$_6$ tetrahedral are considered \cite{ref23}.

To exploit the epitaxial strain modulation on band structures, we conducted magnetoresistance measurements on all SNO/STO samples. Similar as previous the study \cite{ref24}, large linear magnetoresistances (LMR) were observed (see Fig. S4). In addition, Shubnikov de Haas (SdH) oscillations were also observed to be superimposed to the LMR. The oscillation is periodic versus 1/B, and two major frequencies can be discerned (see Figs. 4A\&B). One of them, labeled as $F_1$, rapidly shifts upwards as $d$ increases from 4 to 37 nm, and switches to a slower rise for large $d$ (see Fig. 4E). Another frequency, labeled as $F_2$, starts to be visible only for $d>$ 37 nm and mildly rises with $d$. Intriguingly, $F_1$ again obeys the universal thickness dependence. The coincidence not only unveils the strong modulation of strain on band structure, but also implies that the origin of superconductivity being the associated bands. another weak but discernible frequency ($F_0$) in addition to $F_1$ and $F_2$ exists between 38 and 50 T with no obvious dependence on $d$. Its small magnitude might be due to relatively low mobility of its corresponding Fermi pocket.

Landau-fan diagrams for $F_1$ and $F_2$ are plotted in Figs. 4C\&D for which integer indices of Landau levels are marked at the minima of the oscillating component $\Delta R$ given the presumption of $\rho_{xy}\gg \rho_{xx}$ is satisfied (see Fig. S5). According to Lifshitz-Onsager quantization rules, $S_F\cdot h/eB=2\pi(N+N_0 )=2\pi(N+1/2+\gamma)$, where $S_F$ is the Fermi surface area, $h$ the Planck’s constant, $e$ the electron charge. $N_0=1/2+\gamma$ is the intercept of Landau-fan diagram and $\gamma$ is associated with dimensionality and Berry phase. One can see that $N_0$ of the $F_1$ Landau-fan diagram is a continuous function of $d$, and strikingly, it also obeys the universal thickness dependence.

\begin{figure*}
	\centering
	\includegraphics[width=1\linewidth]{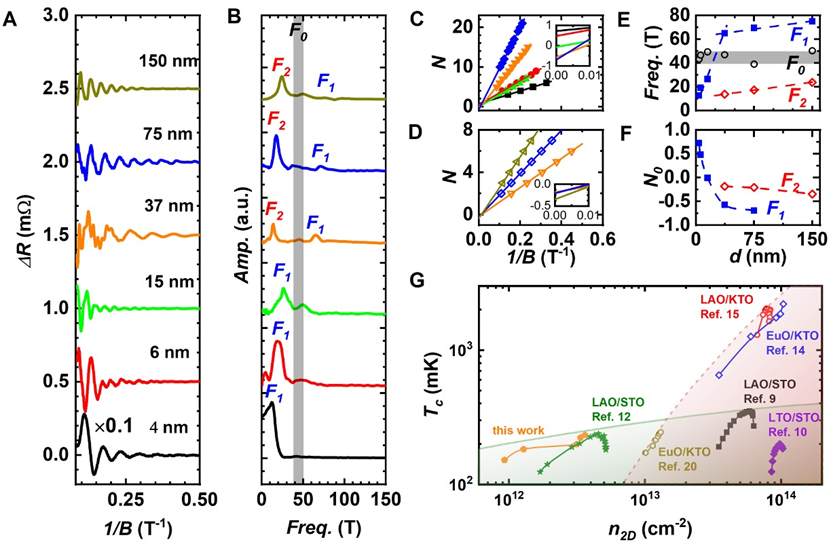}
	\caption{\textbf{(A)} SdH oscillations and \textbf{(B)} Fast Fourier transformation (FFT) spectra for all SNO/STO heterostructures. \textbf{(C)} and \textbf{(D)} are Landau fan diagrams for $F_1$ and $F_2$ oscillations, respectively. The insets show zoom-in views of intercepts of each diagram. \textbf{(E)} Thickness dependence of identified frequencies ($F_0$, $F_1$, and $F_2$). For $d =$ 4, 6, 15 nm, the oscillations are dominated by $F_1$, and for $d =$ 75 and 150 nm, $F_2$ becomes dominant. Clear interference of $F_1$ and $F_2$ can be seen for $d =$ 37 nm, as shown in Panel \textbf{(A)}. The shaded area highlights the frequency range within which a weak spectral peak $F_0$ is observed. \textbf{(F)} Thickness dependence of $N_0$ for both $F_1$ and $F_2$ oscillations. \textbf{(G)} Summary of correlation between $T_c$ and carrier density ($n_{2D}$) for numerous STO and KTO based interfacial superconductivity. Our results conform to the trend and extend the carrier density to a lower record.}
	\label{Fig4}
\end{figure*}

The origin of the superconductivity is the frontmost question to answer for further explorations. As mentioned above, bulk SNO can be convincingly ruled out. Furthermore, interfacial charge-transfer is unlikely neither because polar catastrophe does not exist as SrO, NbO$_2$, and TiO$_2$ layers are all neutral. On the other hand, superconductivity could be introduced to the STO substrate by either metallic interstitial dopants or oxygen vacancies \cite{ref26,ref27,ref28,ref29}. Nb doping is not a plausible reason for our case because HAADF and EELS measurements both confirm the sharpness and cleanness of the interface with negligible chemical intermixing. The absence of superconductivity for the sample with $d=$ 4 nm provides additional support for this conclusion. 

Indeed, oxygen vacancies may possibly be introduced to STO substrates during processes of sample growth \cite{ref26}. Their concentration is associated with both the in-situ osmotic pressure of oxygen and the thickness of the deposited films. It has been shown that a 3 nm thick deposition is enough to saturate the concentration of oxygen vacancies because it prevents further creation of vacancies if deposition continues \cite{ref30}. As a result, oxygen vacancy concentration should already saturate once the deposition of 4 nm thick SNO is completed. In addition, quantitative analyses show that the band carrier density associated with the observed SdH oscillations is $n_{SdH}=2e/h F_1=(0.9\sim 3.6)\times10^{12}$ cm$^{-2}$. If the superconducting electrons are contributed by a typical 5 nm-thick layer doped by oxygen vacancies, the effective three-dimensional carrier density $n_{3D}$ should be $(1.8\sim7.2)\times10^{18} $cm$^{-3}$, below the density threshold for oxygen vacancies to introduce superconductivity in STO ($n_{3D}\approx10^{19}\sim10^{22} $cm$^{-3}$) \cite{ref26,ref27}, not to mention that a $T_c$ higher than 150 mK requires $n_{3D}>10^{20} $cm$^{-3}$ \cite{ref27}. 

One should note that $n_{SdH}$ is 4$\sim $5 orders of magnitude smaller than the carrier density measured by Hall measurements $n_H$ (see Table S1), which represents the total amount of itinerant carriers including those in the bulk SNO. The marked contrast in magnitude implies that the superconducting electron system has a special formation mechanism. Density-functional-theory (DFT) calculations show that a charge accumulation is formed near the interface. Its majority lies on the SNO side (see Fig. S7). The dependence of $n_{SdH} (\propto F_1)$ on d proves that the electron system is from the bulk SNO. Specifically, charges are transferred from bulk SNO towards the interface under a chemical potential gradient. Given the universal thickness dependence shared by strain and $n_{SdH}$, it can be concluded that the chemical potential gradient is highly related with the strain gradient. Such a charge transfer mechanism is novel and completely different from that caused by interfacial polar catastrophe.

The $F_2$ oscillation emerging at 37 nm marks that a second band starts to play a role. Its corresponding band has distinguished feature from that of $F_1$, not only in oscillating frequency but also in effective electron masses which also exhibit similar thickness dependence (see Fig. S6). Its influence on superconductivity is demonstrated by the two-step superconducting transitions as shown in Fig. 2, which is most obvious when $d=$ 37 nm. Furthermore, linear temperature dependence of $H_c$, a feature of two-dimensional superconductivity, is prominent for $d<$ 37 nm. It evolves to be composed of two linear sections with smaller slope for $T<100$ mK for the 75 and 150 nm samples, suggesting interference between the two bands. The bending at low temperatures might suggest that the $F_2$ band introduces some mechanism to cause phase decoherence, effectively lowering its critical field and deviating from the universal dependence.

The heterostructure of SNO/STO is proved as an appealing platform for further exploration because of several unique features. Firstly, the superconductivity was realized with a charge carrier density as low as $0.93\times10^{12} $cm$^{-2}$. It fits the general trend of $T_c$ vs. 2D carrier density ($n_{2D}$) for the STO-based interfacial superconductivity, but extends the lower density limit by a factor of two to our best knowledge (see Fig. 4G). Secondly, the continuous variation of $N_0$ suggests the existence of non-trivial Berry phase and its evolution with strain. Further systematic studies are necessary to explore the band structure’s topology, such as angular resolved photo emission spectroscopy (ARPES), scanning tunneling spectroscopy (STS), and phase-sensitive quantum transport experiments. Nevertheless, it is interesting in itself to question how such a quasi-2D system can be well protected from the metallic bulk SNO to retain its high mobility and to remain in deep quantum extreme.

\begin{acknowledgments}
This work is supported by National Key R\&D Program of China Grant (Grant No. 2021YFA1401902, 2018YFA0305604, 2020YFA0309100), National Natural Science Foundation of China (NSFC) (No. 11874403, 11974390, 12174428), Key Research Program of Frontier Sciences, CAS, (Grant No. ZDBS-LY-SLH0010), Beijing Natural Science Foundation (Grant No. JQ21002), Guangdong-Hong Kong-Macao Joint Laboratory for Neutron Scattering Science and Technology, and the Strategic Priority Research Program (B) of the Chinese Academy of Sciences (Grant No. XDB33030200).


\end{acknowledgments}

\end{document}


	
\title{{\small Supplementary Materials for}\protect\\Dimensionality-confined superconductivity within SrNbO$_3$-SrTiO$_3$ heterostructures}


\maketitle 




\subsection{1.Syntheses SrNbO$_3$ films and heterostructures}
Polycrystalline Sr$_2$Nb$_2$O$_7$ target was prepared by sintering mixtures of SrCO$_3$ and Nb$_2$O$_5$ powder at \SI{1150}{\degreeCelsius} for 24 hours when pressurized to a ceramic pellet with a diameter of 25.4 mm. Stoichiometric SrNbO$_3$ (SNO) films were fabricated on (001)-oriented TiO2-terminated SrTiO$_3$ (STO) and AlO$_2$-terminated LaAlO$_3$ (LAO) substrates by pulsed laser deposition (PLD). During the film growths, the substrate temperature was kept at \SI{750}{\degreeCelsius}, the laser fluence was fixed at $\sim$1.5 J/cm$^2$, and the repetition of laser pulse was at 5 Hz. The samples were grown under a base pressure of $1\times10^{-8}$ Torr and maintained in vacuum while cooling down to room temperature. SNO films with a thickness ranging from 4 to 150 nm were deposited by counting number of laser pulses, and were capped with an ultrathin STO layers (about 5 unit-cells) to prevent the formation of secondary phases due to oxidization of Nb ions. X-ray reflectivity measurements were performed to confirm the layer thickness and interface/surface roughness. The crystallinity of SNO films was checked by XRD $\uptheta-2\uptheta$ scans (see Fig. S1) and asymmetric reciprocal space mappings (RSMs) (see Fig. S2). Cross-sectional transmission electron microscopy measurements were performed on an ARM 200CF (JEOL, Tokyo, Japan) transmission electron microscope operated at 200 kV and equipped with double spherical aberration (Cs) correctors. High-angle annular dark-field (HAADF) images were acquired at acceptance angle of 90–370 mrad. Atomic-resolved electron-energy loss spectra (EELS) were collected at both Ti L- and Nb L-edges, yielding to a sharp interface with negligible chemical intermixing between the SNO films and the STO or LAO substrates. 
\subsection{2.Strain relaxation}
Figure S2A clearly shows the alignment of the intensity peak along $q_x$ for the SNO layer and STO substrate. The alignment is the evidence of the coherent growth of the 4 nm thick SNO on the STO substrate for which the SNO layer is fully strained. Mis-alignment ($\Delta q_x$) of the SNO peak emerges as $d$ is larger than 6 nm, meaning the existence of strain relaxation, and rapidly increases before d reaches 37 nm, meaning the relaxation is significant in this range. It is also supported by the fact that the broadness of the SNO peak also increases significantly in this range. Strain relaxation becomes mild for $d>$37 nm, which can be proved by the dependence of $\Delta q_x$ on $d$, and the similar size of the SNO peak broadness as well (see Figs. S2D-F).
\subsection{3.Electric transport measurements}
Electric transport measurements were conducted mainly in a dilution refrigerator for temperatures lower than 4 K. The temperature dependence of resistance from 300 K to 1.5 K was measured in a physical property measurement system (PPMS). Electric contacts were made at corners of the square-shaped samples by pressing indium. The measurements were performed in van der Pauw (vdP) methods using the combination of Keithley 6221/2182 in “Delta mode”, for which direction of a dc current was periodically switched and voltage was measured 0.2 sec after the switch. Resistance is calculated by the differentiated voltage and current. The “Delta” mode eliminates the thermoelectric potential difference along wirings that may influence the measurements. The current $I=10 \sim 30 \mu$A for measuring superconducting properties and $I=100\sim300 \mu$A for measuring quantum oscillations.
\subsection{4.Calculations of charge density difference near the SNO/STO interface}
Our density functional theory (DFT) calculation is performed by Vienna ab initio simulation package (VASP) code \cite{ref1} with the projector augmented wave (PAW) method \cite{ref2}. The Perdew-Burke-Ernzerhof (PBE) \cite{ref3} exchange-correlation functional is used in our calculation. The kinetic energy cutoff is set to be 500 eV for the expanding the wave functions into a plane-wave basis. The energy convergence criterion is $10^{-8}$ eV and the Gamma-centered k-mesh is used in KPOINTS files. The structures are fully relaxed with built-in 25 Å thick vacuum layer while forces are minimized to less than 0.01 eV/Å. The planar averaged charge density difference (CDD) is calculated by integrating the CDD in the ab plane. It is the difference of charge distribution compared with the bulk phase, which represents the charge accumulation associated with the interface and strain.
Based on the calculations, we can see clear charge accumulations at the interface. As shown in Figure S6, the distribution of the electronic system extends about 2 nm into the SNO layer, and less than 0.5 nm into the STO substrate, meaning its majority lies on the SNO side. It confirms that the electronic system is mainly contributed by charge transfer within the SNO layer.

\clearpage

%
%

\begin{figure}[htb]
	\centering
	\includegraphics{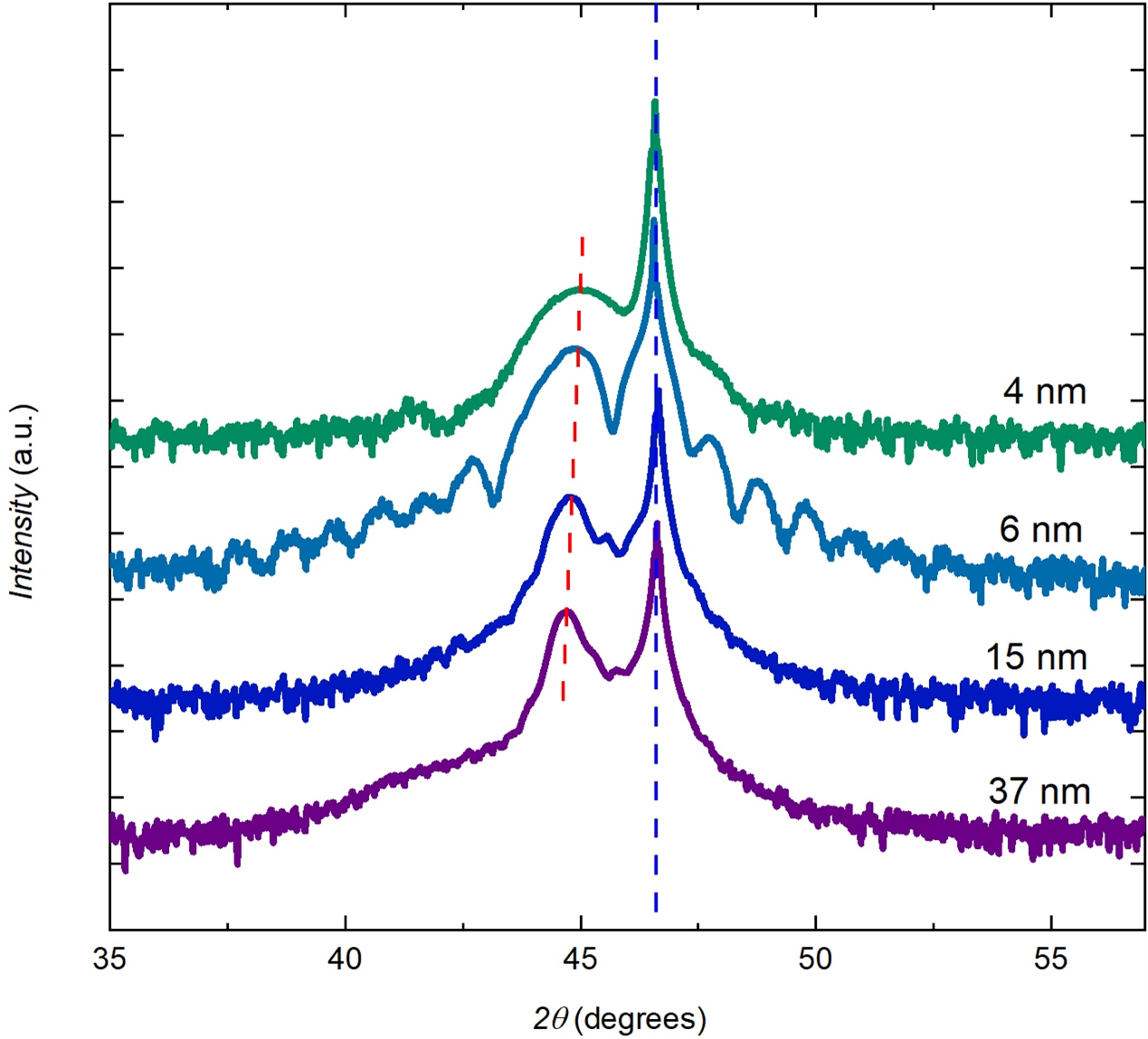}
	\caption{X-ray diffraction (XRD) scan of SNO/STO samples with $d=$ 4,6,15,37 nm. The blue and red dashed lines depict the trajectory of peak evolutions as d increases for STO and SNO, respectively.}
	\label{FigS1}
\end{figure}

\begin{figure}
	\centering
	\includegraphics{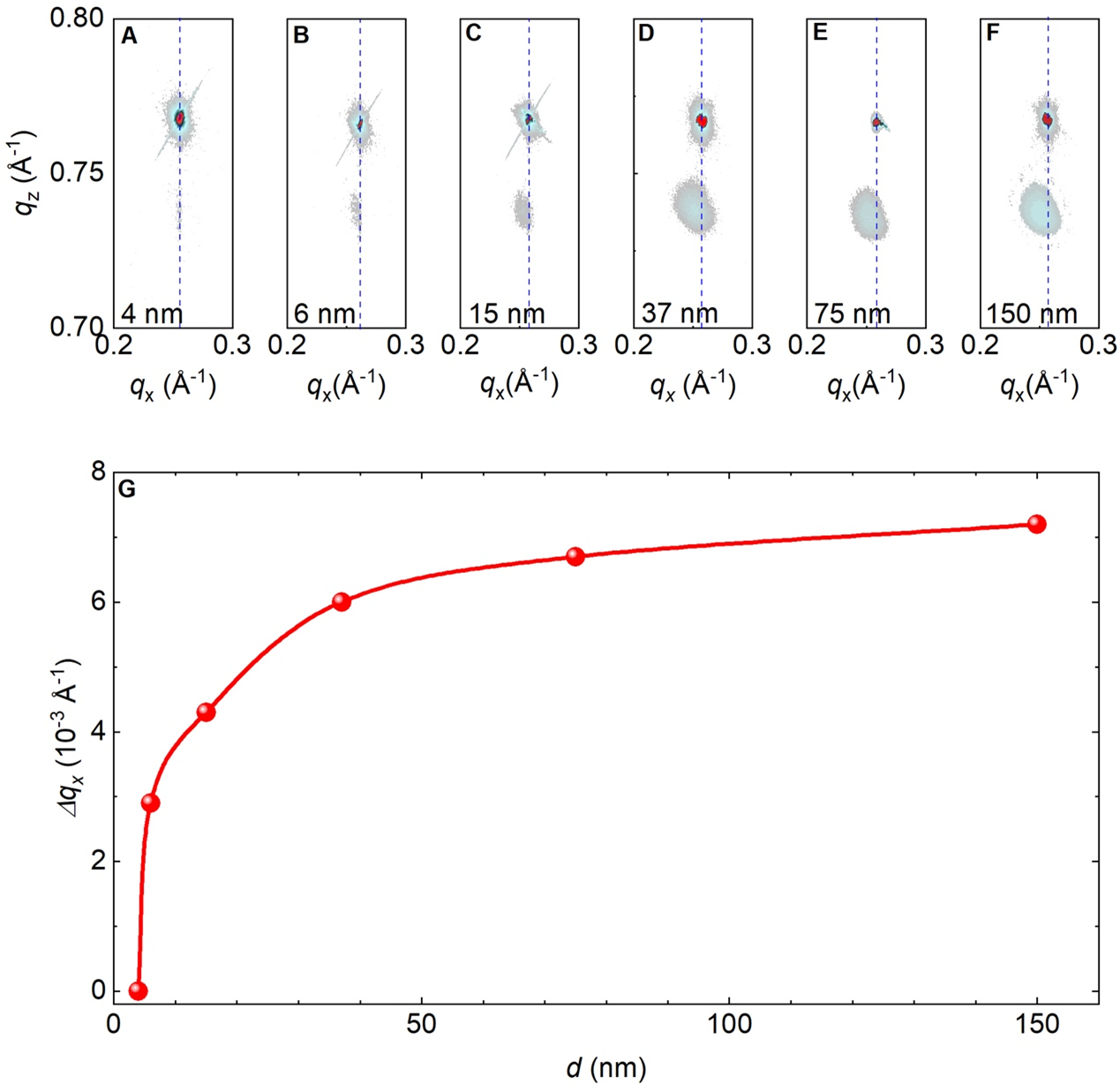}
	\caption{(A-F) RSM results of all samples. RSM peaks contributed by the SNO layers and the STO substrates can be clearly observed. (G) Thickness dependence of mis-alignment between the SNO and STO peaks in $q_x$.}
	\label{FigS2}
\end{figure}

\begin{figure}
	\centering
	\includegraphics{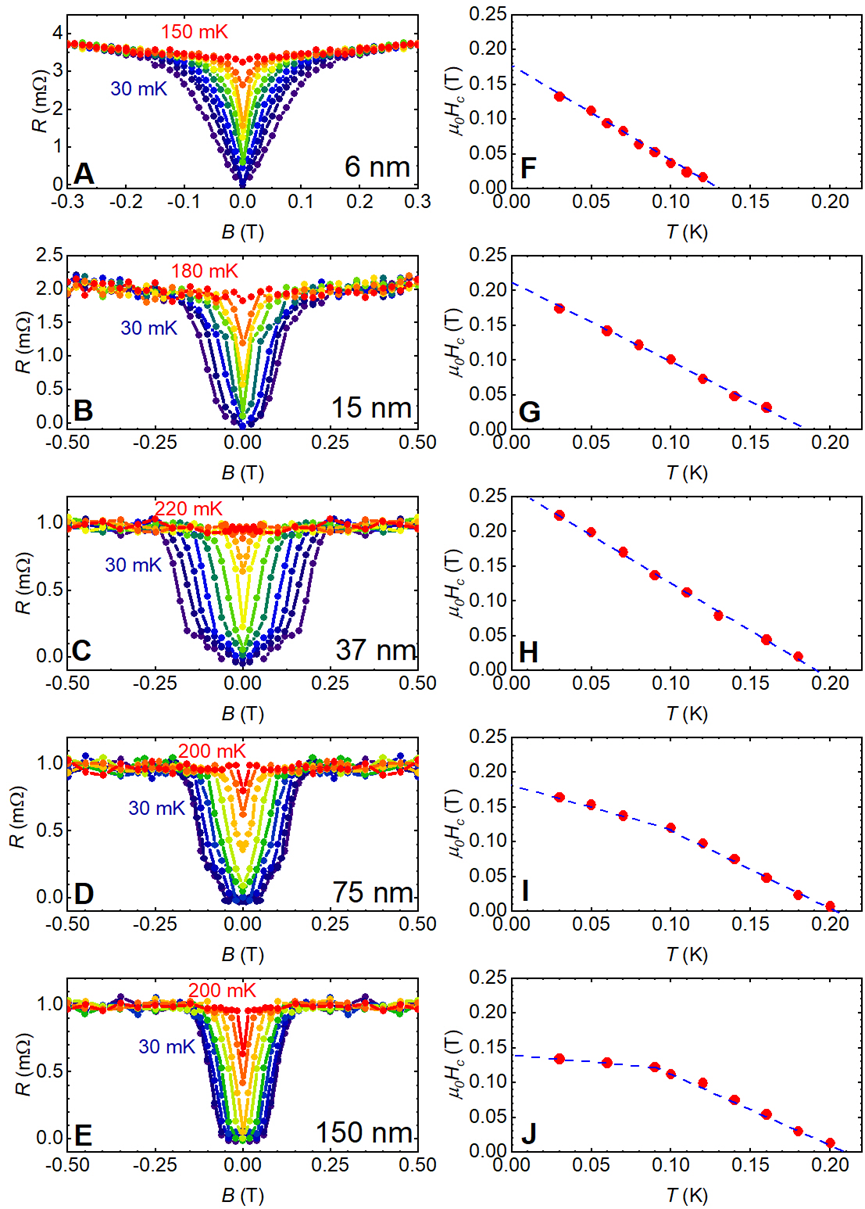}
	\caption{Resistance measured at different temperatures as a function of magnetic field (from A to E) and temperature dependence of critical magnetic field (from F to J) for all superconducting samples.}
	\label{FigS3}
\end{figure}

\begin{figure}
	\centering
	\includegraphics{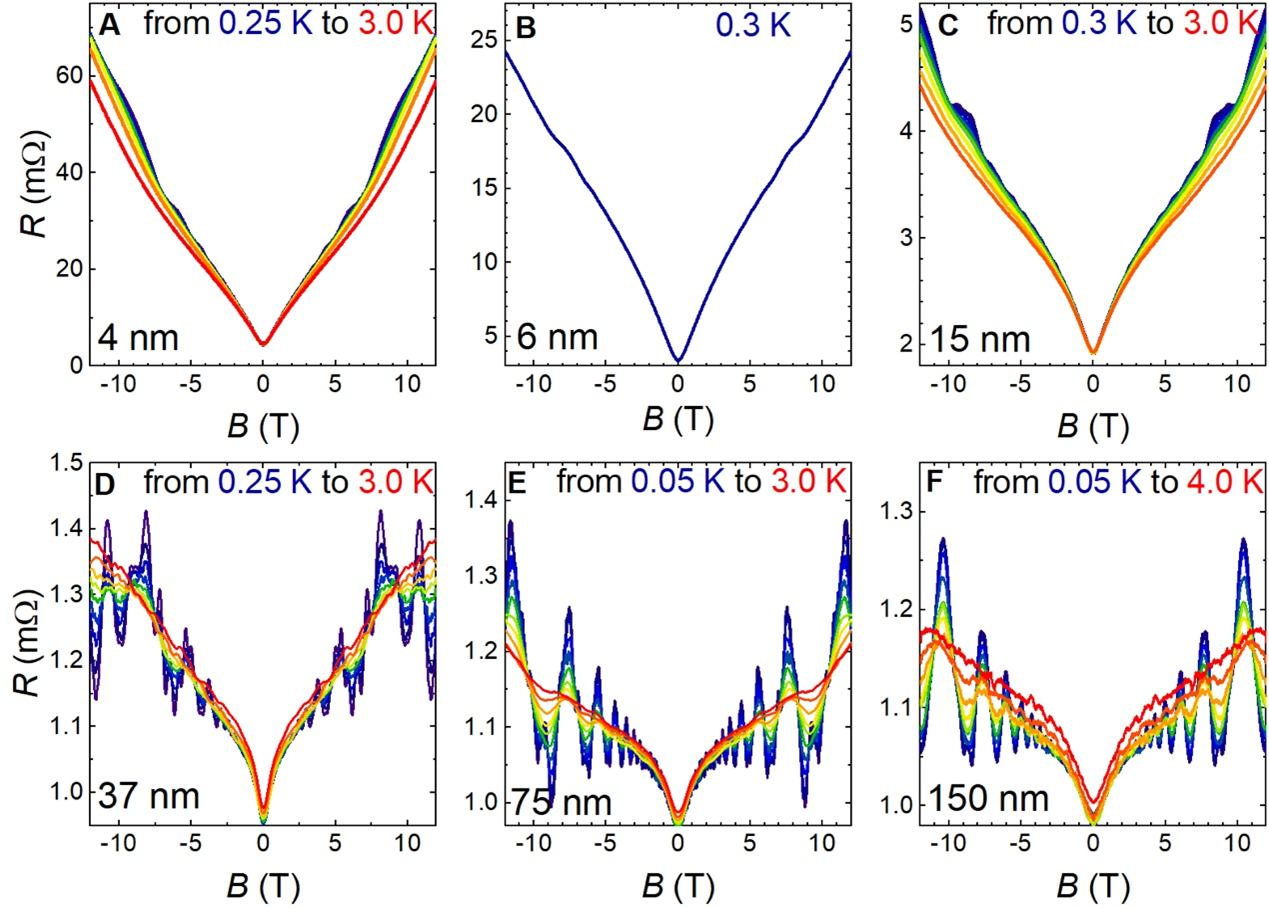}
	\caption{Linear magnetoresistance (LMR) measured at different temperatures all samples. LMR decreases as SNO thickness increases. The change is rapid before d reaches 37 nm, and saturate for larger $d$. LMR of the 6 nm sample was only measured at 0.3 K.}
	\label{FigS4}
\end{figure}

\begin{figure}
	\centering
	\includegraphics{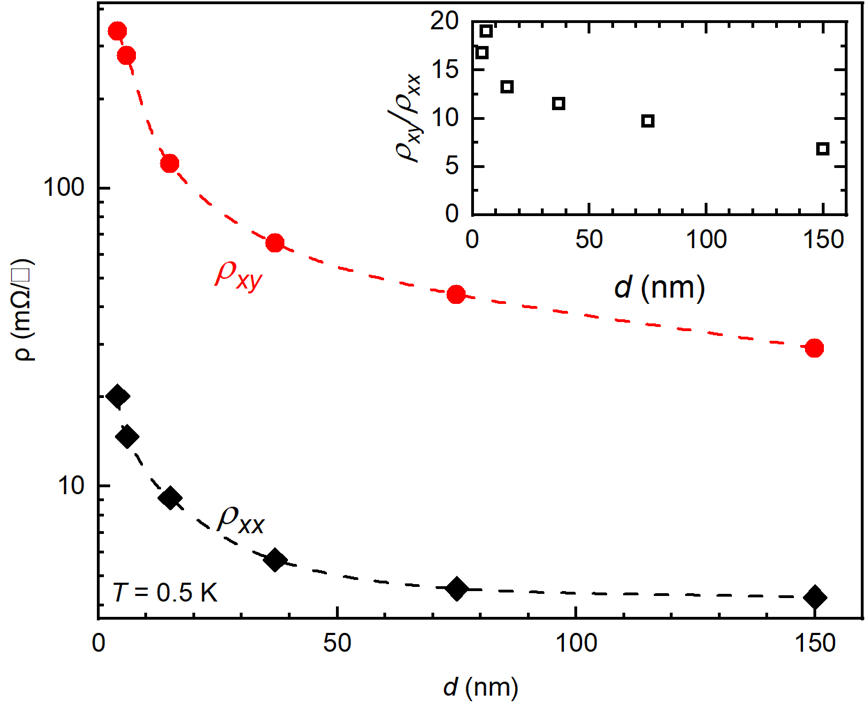}
	\caption{$\rho_{xx}$ and $\rho_{xy}$ as functions of d to prove that the condition of $\rho_{xy}\gg\rho_{xx}$ is always met as shown in the inset. The data are measured at 0.5 K.}
	\label{FigS5}
\end{figure}

\begin{figure}
	\centering
	\includegraphics{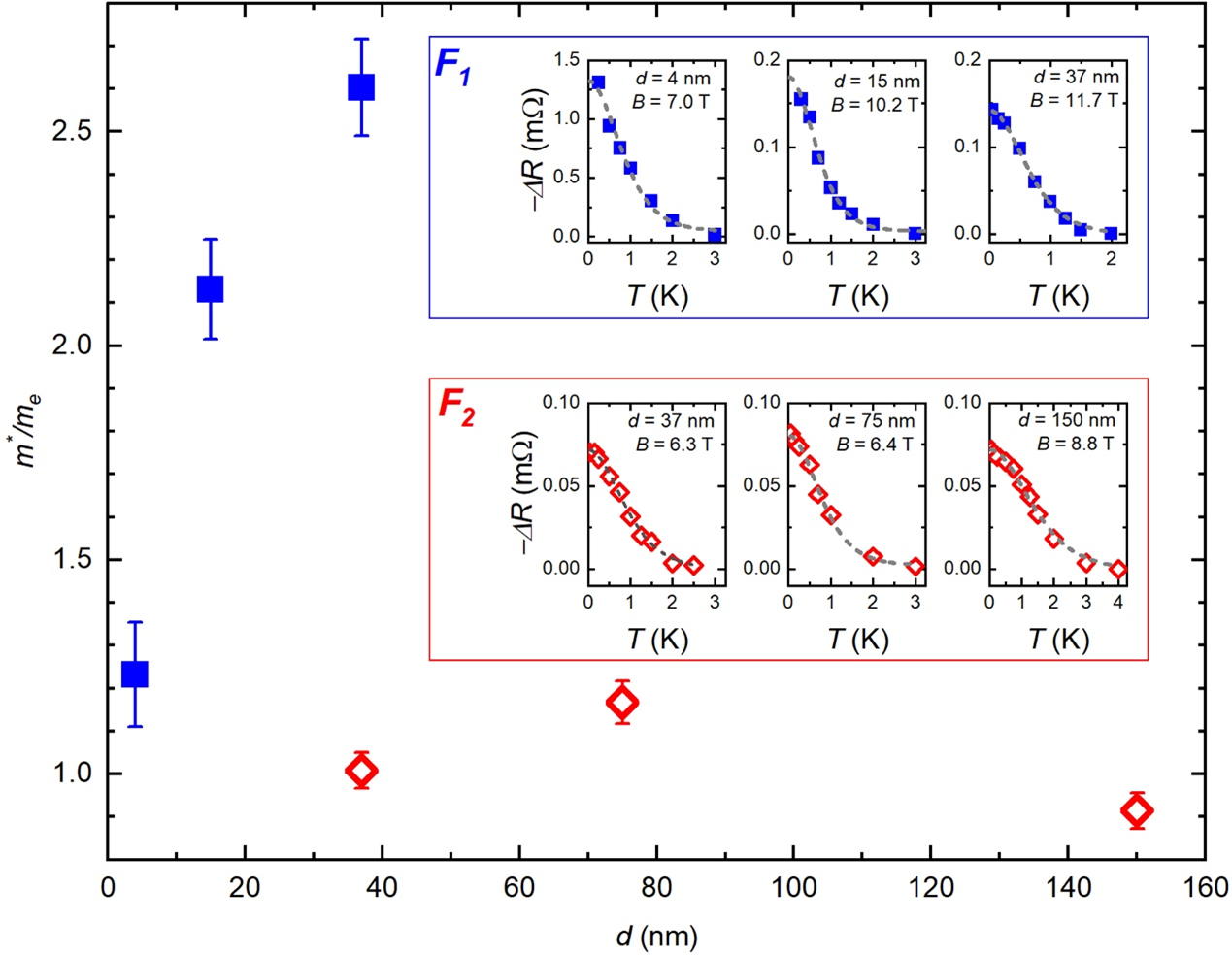}
	\caption{Effective masses extracted from temperature dependence of the SdH oscillations (as shown in the insets). The oscillation amplitudes are read at the valleys where integer indices are labelled. For the 37 nm sample, two amplitudes are simultaneously extracted for both $F_1$ and $F_2$ oscillations since their interference is obvious. The blue solid squares and frame demonstrate the effective mass and temperature dependence of oscillation amplitudes for the $F_1$ oscillations. The red open diamonds and frame demonstrate those for the $F_2$ oscillations.}
	\label{FigS6}
\end{figure}

\begin{figure}
	\centering
	\includegraphics{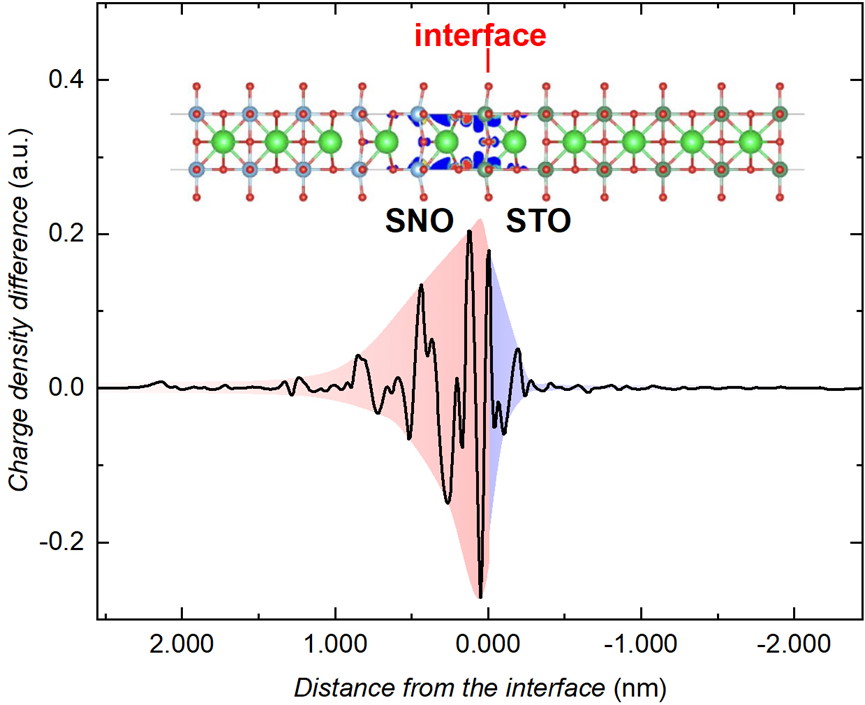}
	\caption{Calculation of charge density difference at the SNO/STO interface compared with bulk phase. The inset figure demonstrates the differential charge density distribution (in blue pockets) near the interface.}
	\label{FigS7}
\end{figure}

	%
		
\clearpage
\begin{table}[htb]
	\caption{Comparison of $n_H$ and $n_{SdH}$.}
	\begin{ruledtabular}
		\begin{tabular}{ccc}
			$d$ (nm)&$n_H$ ($10^{16}$cm$^{-2}$)&$n_{SdH}$ ($10^{12}$cm$^{-2}$)\\
			\hline
			4& 1.98 & 0.618\\
			6& 2.68 & 0.927\\
			15& 6.16 & 1.28\\
			37& 10.9 & 3.14\\
			75& 17.0 & 3.36\\
			150& 25.7 & 3.62\\
		\end{tabular}
	\end{ruledtabular}
\end{table}